\documentstyle[aps,prl,twocolumn,amsfonts,epsfig,floats]{revtex}
\tighten
\newcommand{\mb}[1]{\mbox{\boldmath $#1$}}

\begin{document}
\bibliographystyle{prsty}
\twocolumn[\hsize\textwidth\columnwidth\hsize\csname
@twocolumnfalse\endcsname

\title{String inspired braneworld cosmology}

\author{Cristiano Germani and Carlos F. Sopuerta}

\address{~}

\address{Institute of Cosmology and Gravitation, 
Portsmouth University, Portsmouth~PO1~2EG, United Kingdom}

\address{~}

\date{\today}

\maketitle


\begin{abstract}
We consider braneworld scenarios including the leading correction to the
Einstein-Hilbert action suggested by superstring theory,  
the Gauss-Bonnet term.  
We obtain and study the complete set of equations governing the 
cosmological dynamics.  We find they have the same form as those in 
Randall-Sundrum scenarios but with time-varying four-dimensional 
gravitational and cosmological constants.  Studying the bulk geometry 
we show that this variation is produced by bulk curvature terms 
parameterized by the mass of a black hole.  Finally, we show there is
a coupling between these curvature terms and matter that can be
relevant for early universe cosmology.
\end{abstract}

\vskip 1pc  ]

In recent decades developments in cosmology have been strongly
influenced by high-energy physics.  A remarkable example of this 
is the inflationary scenario and all its variants.
This influence has been growing and becoming more and more important.
Today it comes from developments in
string and M theories and the new scenarios they are providing for
cosmology.  In particular, the study of scenarios where the spacetime
has non-compact extra-dimensions has produced intense activity.  

The  most popular model of that class is the one proposed by Randall and
Sundrum~\cite{RaSu:99a,RaSu:99b} (RS), motivated by orbifold
compactification of higher dimensional string theories, in particular,
by the dimensional reduction of eleven-dimensional supergravity in
${\mathbb{R}}^{10}\times S^1/{\mathbb{Z}}_2$ introduced by
Ho\v{r}ava and Witten~\cite{HoWi}.  The picture
coming from the RS model is one with all matter and gauge fields,
except gravity, confined in a three-brane embedded in a five-dimensional
(5D) spacetime with ${\mathbb{Z}}_2$ symmetry, and where the zero-mode of the
Kaluza-Klein (KK) dimensional reduction is localized, reproducing
Newtonian gravity in the weak field approximation.

On the other hand, it is a general belief that Einstein gravity is a
low-energy limit of a quantum theory of gravity which is still unknown.
Among promising candidates we have string theory which suggests, that
in order to have a ghost-free action, quadratic curvature corrections to 
the Einstein-Hilbert action must be proportional to the Gauss-Bonnet 
term~\cite{BoDe:85}. This term also plays a fundamental role in Chern-Simons 
gravitational theories~\cite{Cha:89}.
However, although being a string-motivated scenario, the RS model and 
its generalizations~\cite{japan} do not include these terms.  
From a geometric point of view, the combination of the 
Einstein-Hilbert and Gauss-Bonnet term constitutes, for 5D spacetimes, 
the most general Lagrangian producing second-order field 
equations~\cite{love1} (see also~\cite{lovelock}).

These facts provide a strong motivation for the study of braneworld
theories including a Gauss-Bonnet term.  
Recent investigations on this issue have shown~\cite{MeOl:00}
that the metric for a vacuum 3-brane (domain wall) is, up to a redefinition
of constants, the warp-factor metric of the RS scenarios. The existence
of a KK zero-mode localized on the
3-brane has also been demonstrated ~\cite{local} (see
also~\cite{Ne:01a,ChNeWe:01}). 
Properties of black hole solutions in 
AdS spacetimes have been studied in~\cite{bhads,Ca:01}.  The cosmological
consequences of these scenarios are less well understood.  This issue has
been studied in~\cite{KiKyLe:99,KiKyLe:00} in the case of a 2-brane
model~\cite{RaSu:99a} and in~\cite{AbMo:01} for a single-brane 
model~\cite{RaSu:99b}. However, in both cases only simple ans\"atze for
the 5D metric (written in Gaussian coordinates as
in~\cite{french})  were considered, e.g. the separability of the
metric components in the time and extra-dimension coordinates.
One can see that this assumption is too strong even in 
RS cosmological scenarios~\cite{french}, where they lead to a
very restrictive class of cosmological models, not representative of
the true dynamics.
Other works with higher-curvature terms in braneworld scenarios
are considered in~\cite{others}.

In this paper we obtain the equations governing the dynamics of 
Friedmann-Robertson-Walker (FRW) cosmological models in braneworld 
theories with a Gauss-Bonnet term.  This includes the derivation of 
the appropriate junction conditions for this kind of theory.  
Then, we investigate the new dynamical cosmological behaviour 
and how it is related to the geometry of the bulk.  

Our starting point is the following action 
\begin{eqnarray}
S & = & \frac{1}{\kappa^2}
\int\! d^5x\sqrt{-g}\left( {\cal L}_{EH}+\frac{\alpha}{2}
{\cal L}_{GB}\right)  \nonumber\\
& & \hspace{2cm} 
+\int\! d^4x\sqrt{-\tilde{g}}\,\left( {\cal L}_m -
2\lambda \right)
\,, \label{action}
\end{eqnarray}
where ${\cal L}_{EH}=R-2\Lambda$ is the Einstein-Hilbert Lagrangian
with a negative cosmological constant, $\Lambda<0$, ${\cal L}_{GB}$
is the Gauss-Bonnet correction
\begin{equation}
{\cal L}_{GB} = R^2-4R^{AB}R_{AB}+R^{ABCD}R_{ABCD}\,,
\end{equation}
and matter fields (${\cal L}_m$) are confined to a 3-brane with
${\mathbb{Z}}_2$ symmetry. Moreover, $\lambda$ is a constant
that coincides with the brane tension in the limit 
${\cal L}_m=0=\alpha$.
Objects corresponding to the brane are written with a tilde
to be distinguished from 5D objects; $g$ is
the metric determinant and $R$, $R_{AB}$, and $R_{ABCD}$ are the
scalar curvature and Ricci and Riemann tensors respectively;
$\kappa$ is the 5D gravitational constant and physical units
in which $c=1$ are assumed.
The sign of the fundamental constant $\alpha$ must be positive
according to the expansions carried out in string theory~\cite{BoDe:85}.
If we write the unit normal to the 3-brane as $\mb{n}=\mb{dw}$ 
(Gaussian coordinates), the field equations are
\begin{eqnarray}
G_{AB}+\Lambda g_{AB}+\alpha H_{AB} = \kappa^2\left[-\lambda
\tilde{g}_{AB}+T_{AB}\right]\delta(w) \,, \label{fieldeq}
\end{eqnarray}
where $H_{AB}$ is the second-order Lovelock tensor~\cite{love1}
\begin{eqnarray}
H_{AB} & = & RR_{AB}-2R_A{}^CR_{BC}-2R^{CD}R_{ACBD} \nonumber \\
& & +R_A{}^{CDE}R_{BCDE}-\textstyle{1\over4}g_{AB}{\cal L}_{GB}
\,, \nonumber
\end{eqnarray}
$T_{AB}$ is the energy-momentum tensor describing the
matter confined on the 3-brane
($T_{AB}n^B=0=\tilde{g}_{AB}n^B$).
Since we are interested in the cosmological consequences of the
theory we take $T_{AB}$ to be of the perfect-fluid type
\begin{eqnarray}
T_{AB} = (\rho+p)u_A u_B + p\tilde{g}_{AB}
\,,\nonumber
\end{eqnarray}
with $u^A$, $\rho$, and $p$, being the fluid velocity ($u^A
u_A=-1$), energy density and pressure respectively.

Homogeneous and isotropic cosmological models can be described 
by the 5D line element~\cite{french} ($i,j,\ldots = 1,2,3$)
\begin{eqnarray}
ds^2 = -n^2(t,y) dt^2 + a^2(t,y)h_{ij}dx^i dx^j + b^2(t,y) dy^2\,,
\label{line}
\end{eqnarray}
where $y$ is the fifth dimension coordinate (the brane is located
at $y=0$) and $h_{ij}$ is a
3-dimensional maximally symmetric metric for the surfaces
$\{t={\rm const.},y={\rm const.}\}$, whose spatial curvature is
parameterized by $k=-1,0,1\,.$ Every hypersurface $y={\rm const.}$
has the metric of a FRW cosmological model.

The first step to solve the field equations~(\ref{fieldeq}) is
to study them in the bulk ($y\neq 0$).  Following~\cite{french}
we have found that a set of functions $\{a(t,y),b(t,y),n(t,y)\}$
constitute a solution of the field equations in the bulk provided 
the following equations are satisfied~\cite{NOTA1}
\begin{eqnarray}
\frac{\dot{a}}{a}\frac{n'}{n}+\frac{a'}{a}\frac{\dot{b}}{b}-
\frac{\dot{a}'}{a} = 0\,, \label{feone}
\end{eqnarray}
\begin{eqnarray}
\Phi + \alpha\Phi^2 = \frac{\Lambda}{6} + \frac{C}{a^4} \,, \label{joly}
\end{eqnarray}
where $C$ is an integration constant and
\begin{eqnarray}
\Phi(t,y) = \frac{1}{n^2}\frac{\dot{a}^2}{a^2} -
\frac{1}{b^2}\frac{a'^2}{a^2}+\frac{k}{a^2} \,. \nonumber 
\end{eqnarray}
The simplicity of equation~(\ref{joly}) is remarkable and so is the
fact that the Einstein-Hilbert and Gauss-Bonnet terms depend on 
the same argument, namely $\Phi\,.$
It would be interesting to know whether or not this happens also
for higher-order terms in $D>5$ spacetimes, i.e., whether the
dependence of the nth-order Lovelock term is of the form $\Phi^n$.

To find the equations for the 3-brane we need to study the junction 
conditions of our theory, which will be different from those
in Einstein gravity~\cite{DeDo:00}. We know that the 5D
metric must be continuous across the brane and that there will be
jumps in its normal derivatives due to the energy-momentum distribution 
on the brane [see Eq.~(\ref{fieldeq})].  The
${\mathbb{Z}}_2$ symmetry makes the metric invariant under the
transformation $y\rightarrow-y$ and hence any metric component
can be written as $A(t,y)=\bar{A}(t,|y|)$.   Therefore, the normal 
derivatives are given by 
\begin{eqnarray}
A'(t,y) = \partial_{|y|}\bar{A}(t,|y|) \mbox{sign}(y) \equiv
\bar{A}_1(t,|y|) \mbox{sign}(y) \,, \nonumber 
\end{eqnarray}
\begin{eqnarray}
A''(t,y) = \partial_{|y|}\bar{A}_1(t,|y|)
+2\bar{A}_1(t,|y|)\delta(y) \,. \nonumber 
\end{eqnarray}
The first derivative has a finite discontinuity across the
brane but its square is continuous by virtue of the ${\mathbb{Z}}_2$
symmetry.  
The second derivative has a non-distributional part, the first
term, and a distributional part whose coefficient gives the
jump across the brane, namely $[A']= A'(t,0^+) - A'(t,0^-) =
2\bar{A}_1(t,0)\,.$   The value of this jump can be obtained by
integrating the tt-component of the field equations~(\ref{fieldeq})
across the brane, that is ($\delta(w)=b^{-1}_o\delta(y)$)
\begin{eqnarray}
\lim_{\epsilon\rightarrow 0}\int^\epsilon_{-\epsilon} dy
\left( G_{tt}+\Lambda g_{tt}+\alpha H_{tt}\right) =
\kappa^2(\rho+\lambda)\frac{n^2_o}{b_o} \,, \nonumber
\end{eqnarray}
where from now on the subscript ``$o$'' will denote the value
of the corresponding quantity on the brane.  Then, the equation
for the jump $\bar{a}_1(t,0)$ is
\begin{eqnarray}
\left[1+2\alpha\left(\frac{1}{n^2_o}\frac{\dot{a}^2_o}{a^2_o}-
\frac{1}{b^2_o}\frac{\bar{a}^2_1}{a^2_o}+\frac{k}{a^2_o}\right)\right]
\frac{\bar{a}_1}{a_ob_o}=-\frac{1}{6}(\lambda+\rho)\kappa^2\,.
\nonumber
\end{eqnarray}
This is a cubic equation for the discontinuity $\bar{a}_1$ which,
for sufficiently small $\alpha$, has only one real solution, 
the other two being complex.  Therefore,
if we require our cosmological equations to have the right
$\alpha\rightarrow 0$ limit we are left with only one solution.  
Introducing it into equation~(\ref{joly}) will lead to the
Friedmann equation for the brane.  Equation~(\ref{joly})
is a quadratic equation in the square of the Hubble function 
$H=\dot{a}_o/(n_oa_o)$, but only one solution has the correct limit 
$\alpha\rightarrow 0$.   Remarkably, the Friedmann equation we
get from that solution can be written in the same form as in the 
RS cosmological scenario (compare with the generalized Friedmann 
equations in~\cite{Ma:00,CaSo})
\begin{eqnarray}
H^2 = \frac{1}{3}\tilde{\kappa}^2_\ast \rho\left(1+\frac{\rho}{2\lambda}
\right)-\frac{k}{a^2_o}+\frac{1}{3}\tilde{\Lambda}_\ast\,,
\label{gfeq}
\end{eqnarray}
where the 4D gravitational coupling and cosmological constants,
$\tilde{\kappa}_\ast$ and $\tilde{\Lambda}_\ast$, are now time
dependent.  The only difference with the corresponding RS equation 
is that here we do not have explicitly the {\em dark radiation} 
term~\cite{Ma:00} proportional to $a^{-4}_o$.  
It is actually included, in a non-linear
way, in $\tilde{\Lambda}_\ast$ [see Eq.~(\ref{fdef}) below]. 
The time-dependent fundamental ``constants'' $\tilde{\kappa}^2_\ast$ 
and $\tilde{\Lambda}_\ast$ change in time as functions only of 
the scale factor $a_o$.  Their explicit form is
\begin{eqnarray}\label{definitions}
\tilde{\kappa}^2_\ast & = & \tilde{\kappa}^2\frac{1+\textstyle{2\over3}
\alpha\Lambda}{1+4\alpha\sigma}\,,~~\tilde{\kappa}^2 \equiv \frac{\lambda
\kappa^4}{6}\frac{1}{1+\textstyle{2\over3}\alpha\Lambda}\,,~~
\sigma \equiv \frac{\Lambda}{6}+\frac{C}{a^4_o}\,, \nonumber \\
\tilde{\Lambda}_\ast & = & \frac{\lambda\tilde{\kappa}^2}{2}\frac{1+
\textstyle{2\over3}\alpha\Lambda}{1+4\alpha\sigma}-
\frac{3}{2\alpha}\left(1-\sqrt{1+4\alpha\sigma}\right) \,, \label{fdef}
\end{eqnarray}
where we have introduced $\sigma$, the function containing the
dependence on $a_o$.  Moreover, $\tilde{\kappa}^2$
is the 4D gravitational coupling constant, the one that
appears in the computation of Newton's law~\cite{Ne:01a}.  
For small $\alpha$ we recover the Friedmann equation in RS braneworlds 
(see, e.g.,~\cite{Ma:00,CaSo}).  Actually, at zero order in $\alpha$,
$\tilde\Lambda_\ast = \tilde{\Lambda}_{RS} + 3Ca^{-4}_o + O(\alpha)\,,$
recovering the dark radiation term $Ca^{-4}_o$ in the Friedmann equation
of RS brane-worlds.

\begin{figure} [t]
\begin{center}
\includegraphics[height=4cm,width=7cm]{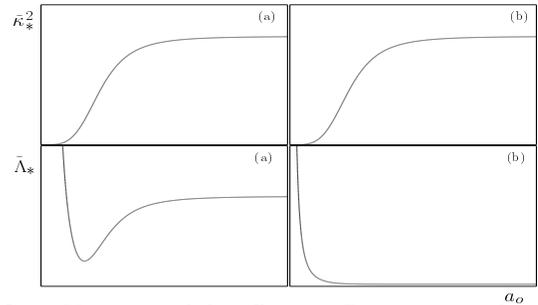} 
\caption{Variation of the effective 4D gravitational
and cosmological constants,  $\tilde{\kappa}^2_\ast$ and
$\tilde{\Lambda}_\ast$, with respect to the scale factor $a_o$
[Eq.~(\ref{fdef})].
(a) $\tilde{\Lambda}_\ast$ has a minimum. 
(b) $\tilde{\Lambda}_\ast$ is a monotonic function of $a_o$.}
\label{plot}
\end{center}
\end{figure}

The dependence of $\tilde{\kappa}^2_\ast$ and $\tilde{\Lambda}_\ast$ 
on $a_o$ has been plotted in Fig.~\ref{plot}. 
There are two possible behaviours according to whether
$\tilde{\Lambda}_\ast$ has a minimum [Fig.~\ref{plot}(a)]
or not [Fig.~\ref{plot}(b)].  In both cases $\tilde{\Lambda}_\ast$
tends to infinity for small $a_o$ and to a constant
\begin{eqnarray}
\tilde{\Lambda}_\infty=\textstyle{{\lambda\tilde{\kappa}^2}\over2}
-\textstyle{3\over{2\alpha}}
(1-\sqrt{1+\textstyle{2\over3}\alpha\Lambda})\,, \nonumber
\end{eqnarray}
when $a_o\rightarrow\infty$.  Then, $\tilde{\Lambda}_\ast$ tends 
to zero for the following critical value of $\lambda$ 
\begin{eqnarray}
\lambda_c=\textstyle{3\over{\alpha\tilde{\kappa}^2}}
(1-\sqrt{1+\frac{2}{3}\alpha\Lambda}) \,. \nonumber
\end{eqnarray}
The case $\lambda=\lambda_c$, $\rho=C=k=0$ corresponds to a
Minkowskian brane~\cite{RaSu:99a,RaSu:99b} (see~\cite{MeOl:00}).
From Eq.~(\ref{definitions}) we must have $1+4\alpha\sigma > 0$, 
hence from the behaviour for big and small $a_o$ we deduce that
$\alpha|\Lambda|\leq 3/2$ and $\alpha C\geq 0$.
On the other hand, the behaviour of $\tilde{\kappa}^2_\ast$,
when $C\neq 0$, is the same independently of the value of the 
parameters; it tends
to zero for small $a_o$ and to $\tilde{\kappa}^2$ [see 
Eq.~(\ref{fdef})] when $a_o\rightarrow\infty\,.$  

As in standard cosmology, the Friedmann equation~(\ref{gfeq}) 
together with the energy-momentum tensor conservation equations 
[a consequence of the divergence-free character of the left-hand 
side of~(\ref{fieldeq})],
\begin{eqnarray}
\dot{\rho} = -3(\rho+p)H \,, \nonumber
\end{eqnarray}
and a barotropic equation of state $p=p(\rho)$, describe completely 
the cosmological dynamics on the brane.

At this point we do not know much about the geometry of the bulk.
For perfect-fluid FRW cosmological models in RS braneworlds, the bulk 
turns out to be the Schwarzschild-AdS 5D 
spacetime~(see~\cite{Kr:99,Id:00}). Since in
our case we have different gravitational equations [Eqs.~(\ref{feone},
\ref{joly})] we do not expect the same bulk.  The knowledge of
the bulk geometry is important to understand the physical
meaning of the integration constant $C$ appearing in the Friedmann
equation~(\ref{gfeq},\ref{fdef}).  Assuming that the fifth
dimension is {\em static}~\cite{french}, $\dot{b}=0$, we have 
found that there is a coordinate change, analogous to the
one used in RS scenarios~\cite{MuShMa}, that brings
the line element~(\ref{line}) to the following static form
\begin{eqnarray}
ds^2 = -f(R)dT^2+\frac{dR^2}{f(R)}+R^2(d\chi^2+\Sigma^2_k(\chi)
d\Omega^2_2) \,,\label{static}
\end{eqnarray}
where $\Sigma_{-1}=\sinh\chi\,,$ $\Sigma_0=\chi\,,$ and
$\Sigma_1=\sin\chi\,,$ and $d\Omega^2_2$ is the unit 2-sphere
metric.  In RS scenarios, Kraus~\cite{Kr:99} and Ida~\cite{Id:00},
found the Schwarzschild-${\rm AdS}_5$ bulk.   When a Gauss-Bonnet
is present, the solution of the field equations~(\ref{fieldeq})
for the line-element~(\ref{static}) is given by (see~\cite{BoDe:85}
for the $k=0$ case and~\cite{Ca:01} for any $k$)
\begin{eqnarray}
f(R) = k +
\frac{R^2}{2\alpha}\left(1-\sqrt{1+\frac{4\kappa^2\alpha M}
{3V_kR^4}+\frac{2}{3}\alpha\Lambda}\right) \,,
\end{eqnarray}
where $V_k$ is the volume of the 3-surface $\{T={\rm const.},R=1\}$.  In
this picture, the 3-brane is a hypersurface that expands 
or contracts according to a law $R=R(T)$, where $R$ coincides with 
the scale factor, $R(T)=a(t,y)$, and $t$ is a proper time ($n_o=1$).   
More important, the constant $C$ is related to the black hole mass 
by the relation
\begin{eqnarray}
C=\frac{\kappa^2 M}{3V_k}\,. \nonumber
\end{eqnarray}
Therefore, the string theory prediction that $\alpha$ must be positive
implies, using $\alpha C\geq 0$, that the mass of the black hole must be
positive.  Conversely,  if we require the avoidance of a naked 
singularity, $M>0$~\cite{Ca:01},  then the constant $\alpha$ must 
be positive.

As one would expect, if the mass of the black hole tends to zero
the bulk Weyl tensor vanishes.  This means that the effective 
time-variation of $\tilde{\kappa}^2_\ast$ is due to the coupling of 
bulk Weyl curvature terms with matter.   From a geometrical point of 
view, it is a consequence of the structure of the curvature quadratic
terms in the Gauss-Bonnet Lagrangian.  

Let us now study what is the behaviour of FRW models in this theory. 
We start with the most simple case, when there is no black hole present
in the bulk, which means $M=0$.  In this case the dynamics is completely
equivalent to that of RS braneworlds, widely discussed in the literature 
(see~\cite{CaSo,SzDaKr:02} for an exhaustive study), where 
$\tilde{\kappa}$ and $\Lambda_\infty$ are the 4D gravitational
and cosmological constants respectively.  Moreover, one can check
that in this case the conditions for inflation are the same as those 
described in~\cite{MaWaBaHe:00}.

To study the case $M\neq 0$ we assume a linear equation of state, 
$p=(\gamma-1)\rho$, which implies $\rho \propto a_o^{-3\gamma}\,.$ 
For late times, $a_o\gg 1$, and as in RS braneworlds, the behaviour
is the same as in standard cosmology, $a_o\approx t^{2/(3\gamma)}$, but
with the modified 4D constants $\tilde{\kappa}$ and $\Lambda_\infty$.
This is expected since it corresponds to the low-energy regime. At
high energies, or early times ($a_o \ll 1$), things are very 
different.   In the case $\gamma\geq 1$ we find that $H^2 \approx 
a^{4-6\gamma}_o$, hence, $a_o \approx t^{1/(3\gamma-2)}$.  The generic 
behaviour for $\gamma<1$ is $H^2 \approx a^{-2}_o$, therefore 
$a_o \approx t \,.$ Then, when $\gamma<1$ the dynamics is 
independent of the particular equation of state.  This results
show that the dynamical behaviour in our theory is completely different
from the general relativistic case ($a_o \approx t^{2/(3\gamma)}$), 
except for the radiation case ($\gamma=4/3$), when they remarkably 
coincide.  It is also different from the behavior in RS braneworlds 
($a_o \approx t^{1/(3\gamma)}$). 

In conclusion, we have studied the cosmological dynamics in
a theory that generalizes the RS scenario~\cite{RaSu:99a,RaSu:99b}  
by taking into account the Gauss-Bonnet higher-order curvature 
term~\cite{BoDe:85,Cha:89}.   Using this fact, we have shown that the 
cosmological equations can be seen as those of RS scenarios but with 
time-dependent 4D gravitational and cosmological constants.  Studying 
the 5D geometry of our model we have found that the time variation of 
the constants is parametrized only by the mass of a black hole in the 
bulk.  In the case of the gravitational constant, the time dependence is 
introduced through the coupling between bulk curvature terms and matter.  
Finally, we have shown how the higher-order curvature terms present in 
our theory are dominant at high energies and change the cosmological 
dynamics at early times. Hence, they can provide alternative cosmological
scenarios for the study of unsolved cosmological problems.

The authors wish to thank Carlos Barcel\'o, Bruce Bassett, Toni Campos, 
Roy Maartens, Martin O'Loughlin and Carlo Ungarelli for very fruitful 
discussions.  CG is supported by a PPARC studentship and thanks SISSA 
for hospitality during the realization of some parts of this work.  
CFS is supported by the European Commission (contract HPMF-CT-1999-00149).



\begin{thebibliography}{10}
\bibitem{NOTA1} For any function $A=A(t,y)$ we use the notation: $\dot{A} = \partial_t A$
and $A' = \partial_y A\,.$
\end{thebibliography}


\end{document}